\documentclass[reprint,
 amsmath,amssymb,
 aps,
]{revtex4-2}

\usepackage{graphicx}
\usepackage{dcolumn}
\usepackage{bm}
\usepackage{xcolor}

\begin{document}


\title{Arbitrarily accelerating space-time wave packets}

\author{Layton A. Hall$^{1}$}
\author{Murat Yessenov$^{1}$}
\author{Ayman F. Abouraddy$^{1,*}$}
\affiliation{$^{1}$CREOL, The College of Optics \& Photonics, University of Central~Florida, Orlando, FL 32816, USA}
\affiliation{$^*$Corresponding author: raddy@creol.ucf.edu}

\begin{abstract}
All known realizations of optical wave packets that accelerate along their propagation axis, such as Airy wave packets in dispersive media or wave-front-modulated X-waves, exhibit a constant acceleration; that is, the group velocity varies linearly with propagation. Here we synthesize space-time wave packets that travel in free space with arbitrary axial acceleration profiles, including group velocities that change with integer or fractional exponents of the distance. Furthermore, we realize a composite acceleration profile: the wave packet first accelerates from an initial to a terminal group velocity, decelerates back to the initial value, and then travels at a fixed group velocity. These never-before-seen optical-acceleration phenomena are all produced using the same experimental arrangement that precisely sculpts the wave packet's spatio-temporal spectral structure. 
\end{abstract}


\maketitle

\begin{figure}[b!]
\centering 
\includegraphics[width=8.6cm]{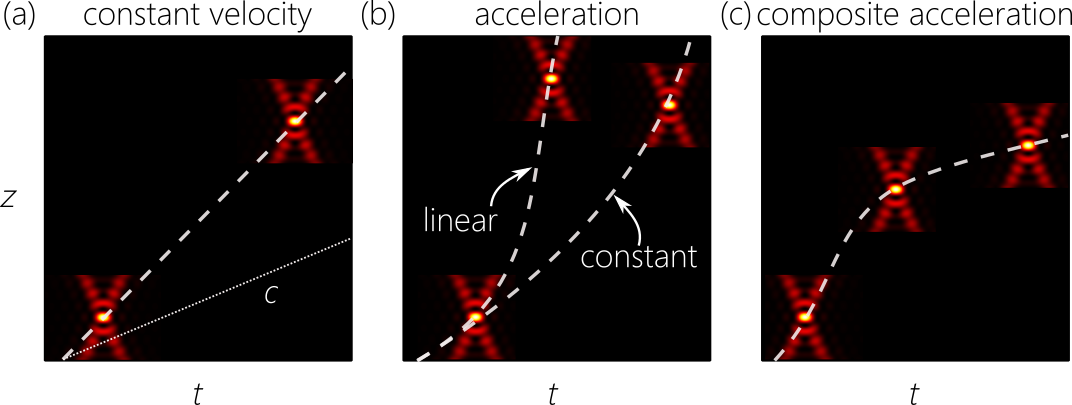} 
\caption{(a) The trajectory of a wave packet in space-time traveling at a constant group velocity $\widetilde{v}$; (b) at a constant acceleration and at a linearly varying rate of acceleration; and (c) with a composite acceleration profile in which acceleration is followed by deceleration. Here $z$ is the axial coordinate, and the wave packet is depicted X-shaped.}
\label{Fig:Illustration}
\end{figure}

In free space, a plane-wave optical pulse travels at a fixed group velocity equal to the speed of light in vacuum $c$. The pioneering work of Berry and Balasz introduced the Airy wave packet \cite{Berry79AMP}: a temporal wave packet satisfying the non-relativistic Schr{\"o}dinger equation that undergoes \textit{constant} acceleration in absence of external forces. Rather than the expected constant-group-velocity propagation [Fig.~\ref{Fig:Illustration}(a)], the group velocity of the Airy wave packet increases linearly along the propagation axis [Fig.~\ref{Fig:Illustration}(b)]. Accelerating \textit{optical} Airy pulses have since been realized in dispersive media, and changes in the group velocity $\widetilde{v}$ of $\Delta\widetilde{v}\!\approx\!10^{-4}c$ over a distance of $\sim\!75$~cm in glass have been reported \cite{Chong10NP} (see also \cite{Abdollahpour10PRL,Efremidis19Optica}). A different approach to achieve acceleration in free space exploits the spatial degree of freedom. Specifically, this can be achieved by precise modulation of the wave front of a propagation-invariant X-wave \cite{Saari97PRL} as proposed by Clerici \textit{et al}. in \cite{Clerici08OE}, or by changing the structure of a cylindrically symmetric tilted pulse front via an axicon \cite{Li20SR}. Experiments have yielded accelerations corresponding to $\Delta\widetilde{v}\!\approx\!10^{-3}c$ and deceleration $\Delta\widetilde{v}\!\approx\!-3\!\times\!10^{-5}c$ over a distance of $\sim\!20$~cm in free space \cite{Lukner09OE}. To date, these have been the only successfully realized accelerating optical wave packets in a homogeneous medium.

Two limiting features emerge: (1) the observed acceleration is small $\Delta\widetilde{v}\!\ll\!c$ (compare to water-wave Airy pulses in which $\Delta\widetilde{v}\!\approx\!0.4\widetilde{v}$ \cite{Fu2015PRL}); and (2) solely \textit{constant} acceleration (linear rate of change in $\widetilde{v}$) has been produced [Fig.~\ref{Fig:Illustration}(b)]. The first limitation stems from the restricted range of values of group velocity typically accessible in optics. We recently \cite{Yessenov20PRL2} addressed this limitation by utilizing propagation-invariant `space-time' (ST) wave packets \cite{Kondakci16OE,Parker16OE,Kondakci17NP,Porras17OL,Efremidis17OL,Yessenov19OPN} that offer group-velocity tunability over an unprecedented span \cite{Kondakci19NC,Bhaduri19Optica,Yessenov19OE,Bhaduri20NP} to demonstrate record-high acceleration approaching $\Delta\widetilde{v}\!\sim\!c$ over a distance of 20~mm, which exceeds previous results by $\sim\!4-5$ orders-of-magnitude \cite{Yessenov20PRL2}. The propagation invariance of ST wave packets \cite{Wong17ACSP2,Kondakci18PRL,Yessenov19OE,Kondakci19NC,Bhaduri19Optica,Bhaduri20NP,Wong20AS} arises from the one-to-one association between their spatial and temporal frequencies \cite{Donnelly93ProcRSLA,Saari04PRE,Longhi04OE,Kondakci19OL,Yessenov19PRA}. The transition to accelerating ST wave packet necessitates associating each wavelength with a finite spatial bandwidth -- rather than a single spatial frequency -- whose center and bandwidth are wavelength dependent.

The restriction to \textit{constant} acceleration has been recently examined theoretically by Li \textit{et al}. who extended their earlier work \cite{Li20SR} by controllably deforming the wave front of a tilted pulse front before an axicon to produce composite acceleration profiles \cite{Li20CP} (related results have been described \cite{Li21CP} for `flying-focus' wave packets \cite{SaintMarie17Optica,Froula18NP,Jolly20OE}). However, the changes in $\widetilde{v}$ remain small for a realistic selection of the parameters involved.

\begin{figure*}[t!]
\centering 
\includegraphics[width=17.6cm]{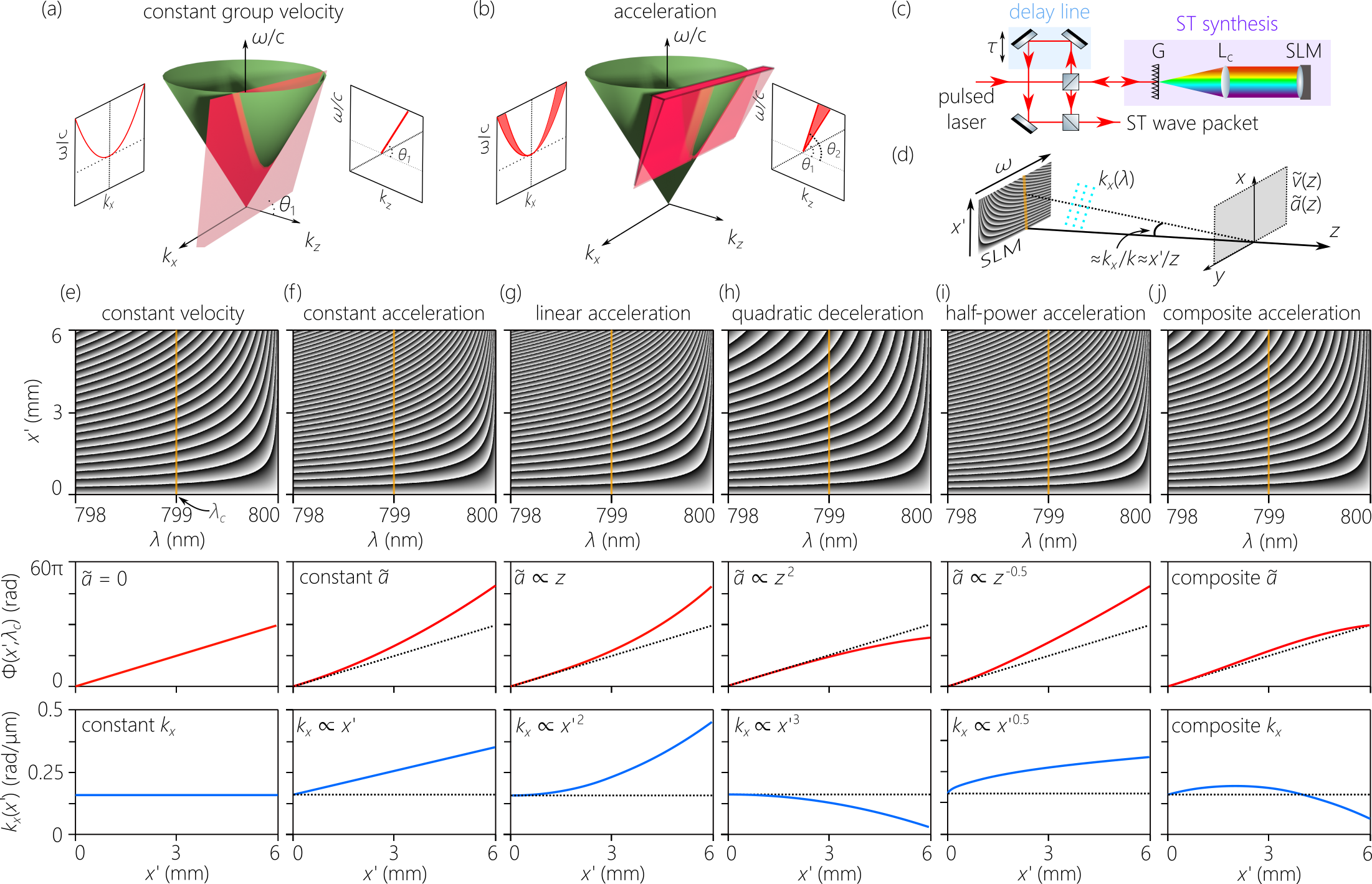} 
\caption{(a) The spectral support domain on the free-space light-cone of a propagation-invariant ST wave packet traveling at a fixed group velocity $\widetilde{v}_{1}\!=\!c\tan{\theta_{1}}$, along with the spectral projections onto the $(k_{x},\tfrac{\omega}{c})$ and $(k_{z},\tfrac{\omega}{c})$ planes. (b) Same as (a) for a ST wave packet accelerating between $\widetilde{v}_{1}\!=\!c\tan{\theta_{1}}$ and $\widetilde{v}_{2}\!=\!c\tan{\theta_{2}}$. (c) Schematic depiction of the setup to synthesize and characterize accelerating ST wave packets. G: Grating, L$_{\mathrm{c}}$: cylindrical lens, SLM: spatial light modulator. (d) Geometry of the modulated field propagating from the SLM to the axial plane $z$. (e-j) In each panel, we provide the 2D SLM phase distribution $\Phi(\lambda,x')$, the phase $\Phi(\lambda_{\mathrm{c}},x')$ at $\lambda_{\mathrm{c}}\!=\!799$~nm corresponding to the vertical line in $\Phi(\lambda,x')$, and its derivative $k_{x}(\lambda_{\mathrm{c}},x')\!=\!\tfrac{d\Phi}{dx'}$ corresponding to the local spatial frequency. The dotted lines correspond to $k_{x}\!=\!0.16$~rad$/\mu$m. (e) Constant-$\widetilde{v}$, $m\!=\!0$; (f) constant-$\widetilde{a}$, $m\!=\!1$; (g) linear acceleration, $m\!=\!2$; (h) quadratic acceleration, $m\!=\!3$; (i) fractional exponent $m\!=\!0.5$; and (j) a composite acceleration profile.}
\label{Fig:ConceptAndSetup}
\end{figure*}

Here we demonstrate for the first time optical wave packets with \textit{arbitrary axial acceleration profiles}. Rather than a constant acceleration, we produce wave packets whose group velocity changes linearly or superlinearly, with integer or fractional exponents [Fig.~\ref{Fig:Illustration}(b)]. Furthermore, we produce a composite acceleration profile, whereby the wave packet first accelerates from an initial group velocity $\widetilde{v}_{1}$ to a terminal value $\widetilde{v}_{2}$, and then decelerates back from $\widetilde{v}_{2}$ to $\widetilde{v}_{1}$ [Fig.~\ref{Fig:Illustration}(c)]. These results are achieved while maintaining the large changes in group velocities from \cite{Yessenov20PRL2}. By establishing a general methodology for sculpting the spatio-temporal spectrum via phase-only modulation, arbitrary acceleration profiles are readily produced.

We first describe the structure of propagation-invariant ST wave packets. Because each spatial frequency $k_{x}$ (the transverse spatial number) is associated with a single temporal frequency $\omega$, the spectral support domain on the surface of the free-space light-cone $k_{x}^{2}+k_{z}^{2}\!=\!(\tfrac{\omega}{c})^{2}$ is a 1D trajectory, where $k_{z}$ is the axial wave number, and we hold the field uniform along $y$ for simplicity \cite{Kondakci17NP}. Propagation invariance requires that this trajectory be a conic section at the intersection of the light-cone with a plane $\Omega\!=\!(k_{z}-k_{\mathrm{o}})c\tan{\theta}$, which is parallel to the $k_{x}$-axis and makes an angle $\theta$ (the spectral tilt angle) with the $k_{z}$-axis; here $\Omega\!=\!\omega-\omega_{\mathrm{o}}$, $\omega_{\mathrm{o}}$ is a carrier frequency, and $k_{\mathrm{o}}\!=\!\tfrac{\omega_{\mathrm{o}}}{c}$ [Fig.~\ref{Fig:ConceptAndSetup}(a)]. The spectral projection onto the $(k_{x},\tfrac{\omega}{c})$-plane can be approximated by a parabola in the vicinity of $k_{x}\!=\!0$: $\tfrac{\Omega}{\omega_{\mathrm{o}}}\!=\!\tfrac{k_{x}^{2}}{2k_{\mathrm{o}}^{2}(1-\widetilde{n})}$; where $\widetilde{n}\!=\!\cot{\theta}$ is an effective group index. The projection onto the $(k_{z},\tfrac{\omega}{c})$-plane is a straight line that makes an angle $\theta$ with the $k_{z}$-axis. The resulting ST wave packet travels rigidly in free space at a fixed group velocity $\widetilde{v}\!=\!c\tan{\theta}\!=\!c/\widetilde{n}$ \cite{Kondakci19NC,Yessenov19OE}. The spectral support for a ST wave packet traveling at a constant acceleration between terminal group velocities $\widetilde{v}_{1}\!=\!c\tan{\theta_{1}}$ and $\widetilde{v}_{2}\!=\!c\tan{\theta_{2}}$ results from the light-cone intersecting with a wedge that is parallel to the $k_{x}$-axis and whose faces makes angles $\theta_{1}$ and $\theta_{2}$ with the $k_{z}$-axis [Fig.~\ref{Fig:ConceptAndSetup}(b)], thereby producing a 2D spectral support domain \cite{Yessenov20PRL2}. We consider a general group-velocity profile $\widetilde{v}(z)\!=\!\widetilde{v}_{1}+\tfrac{1}{m}\Delta\widetilde{v}(\tfrac{z}{L})^{m}$ and associated acceleration profile $\widetilde{a}(z)L\!=\!\Delta\widetilde{v}(\tfrac{z}{L})^{m-1}$, where $\widetilde{a}(z)\!=\!\tfrac{d\widetilde{v}}{dz}$, $L$ is the maximum propagation distance, $\widetilde{v}_{1}\!=\!\widetilde{v}(0)$ is the initial group velocity, $\widetilde{v}_{2}\!=\!\widetilde{v}(L)$ is the terminal group velocity, $\Delta\widetilde{v}\!=\!\widetilde{v}_{2}-\widetilde{v}_{1}$, $m$ is an acceleration exponent that need not be an integer, and $\widetilde{v}(z)\!=\!\widetilde{v}_{1}$ when $m\!=\!0$. To date, only $m\!=\!0$ (constant-$\widetilde{v}$) and $m\!=\!1$ (constant-$\widetilde{a}$) have been realized. 

\begin{figure*}[t!]
\centering 
\includegraphics[width=18.4cm]{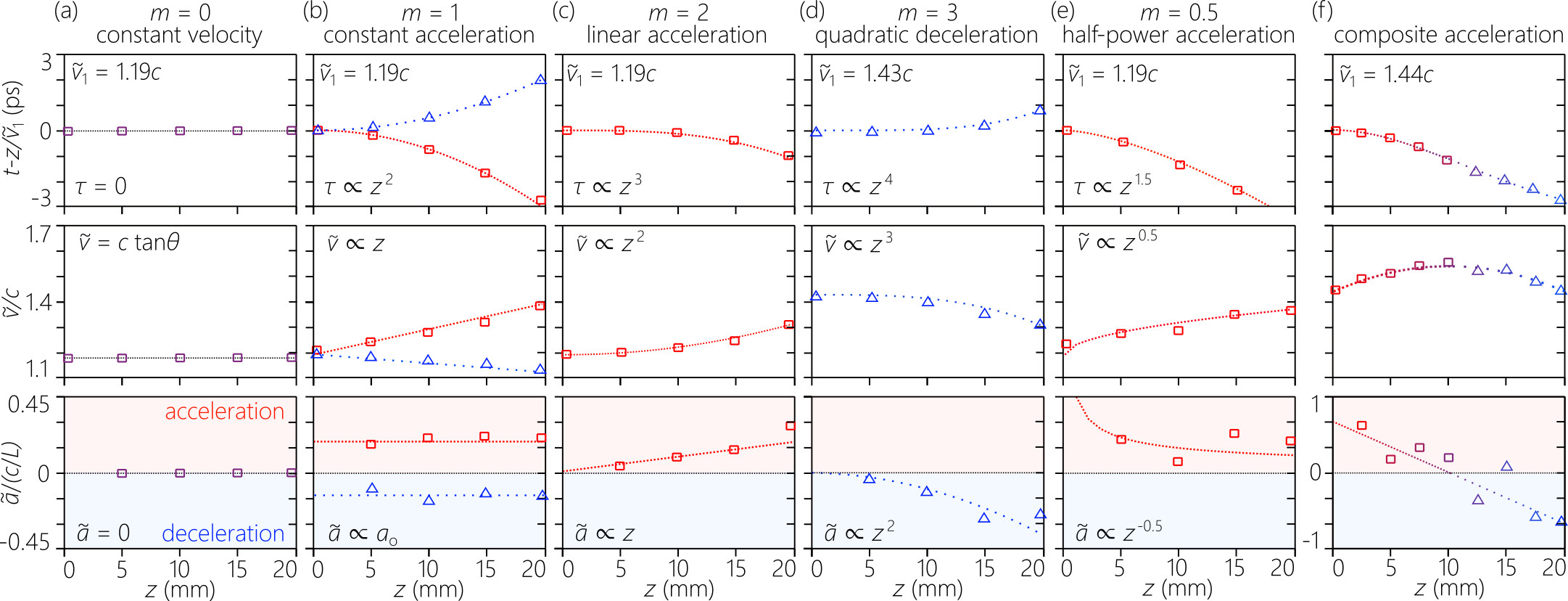} 
\caption{Measurements of arbitrary group-velocity and acceleration profiles. For each case we plot in the first row the group delay measured along the propagation axis $z$ in a frame moving at the initial group velocity $\widetilde{v}_{1}\!=\!\widetilde{v}(0)$, the axial distribution of the group velocity $\widetilde{v}(z)$ in the second row, and the axial distribution of the acceleration $\widetilde{a}(z)$ in the third row. (a) A propagation-invariant ST wave packet with constant group velocity $\widetilde{v}\!=\!1.19c$, corresponding to a spectral tilt angle $\theta\!=\!50^{\circ}$, and $\widetilde{a}(z)\!=\!0$. (b) Two ST wave packets, one propagating at a constant acceleration, and the other at a constant deceleration. (c) A ST wave packet with linearly varying acceleration, $\widetilde{a}(z)\!\propto\!z$. (d) A ST wave packet with quadratic deceleration such that $\widetilde{a}(z)\!\propto\!-z^{2}$. (e) A ST wave packet with fractional acceleration rate $\widetilde{a}(z)\!\propto\!z^{-0.5}$. (f) A ST wave packet with a composite acceleration profile.}
\label{Fig:MeasuredAcceleration}
\end{figure*}

We synthesize ST wave packets using the pulsed-beam shaper in Fig.~\ref{Fig:ConceptAndSetup}(c). A diffraction grating G (1200~lines/mm) and a cylindrical lens L$_{\mathrm{c}}$ collimate the spatially spread spectrum of femtosecond pulses from a Ti:sapphire laser (Tsunami, Spectra Physics; central wavelength $\approx\!800$~nm and pulsewidth $\approx\!100$~fs). A reflective, phase-only spatial light modulator (SLM) placed at the focal plane imparts a 2D phase distribution $\Phi$ to the impinging spectrally resolved wave front. The retro-reflected field is reconstituted after the grating G into a ST wave packet with bandwidth $\Delta\lambda\!\approx\!1$~nm (pulsewidth $\approx\!2$~ps). The axial distribution of the group velocity $\widetilde{v}(z)$ is reconstructed by interfering 100-fs reference pulses from the laser after an optical delay $\tau$ with the ST wave packets. At each axial position, we estimate the required group delay to ensure the overlap of the ST wave packet with the reference pulse in space and time, as evinced by high-visibility spatially resolved interference fringes \cite{Kondakci19NC,Bhaduri19Optica,Bhaduri20NP,Yessenov20PRL2}.

The 2D SLM phase distribution $\Phi$ determines the structure of the synthesized ST wave packet. To produce a propagation-invariant ST wave packet at a group velocity $\widetilde{v}\!=\!c/\widetilde{n}$, $\Phi$ is designed to associate with each frequency $\Omega$ the spatial frequency $ck_{x}(\Omega)\!=\!\sqrt{2\omega_{\mathrm{o}}\Omega(1-\widetilde{n})}$ by imparting a spatial phase $\Phi(\Omega,x')\!=\!k_{x}(\Omega)x'$ that is linear in $x'$ [Fig.~\ref{Fig:ConceptAndSetup}(e)]. Tuning $\widetilde{v}$ is then achieved by modifying $\Phi$ to vary the group index $\widetilde{n}$.

To synthesize \textit{accelerating} ST wave packets, we establish a mapping from the axial group-velocity profile $\widetilde{v}(z)$ to the SLM phase $\Phi(\Omega,x')$ for each $\Omega$ to assign a new spectral tilt angle $\theta(z)$ to each axial plane $z$, $\widetilde{n}(z)\!=\!\cot{\{\theta(z)\}}$. This strategy replaces the single $k_{x}$ assigned to each $\Omega$ with a finite spatial bandwidth endowed with a particular chirp profile matched to $\widetilde{v}(z)$, thus producing in principle an arbitrary acceleration profile $\widetilde{a}(z)$. The SLM phase $\Phi(\Omega,x')$ is no longer linear in $x'$, and the \textit{local} spatial frequency is $k_{x}(\Omega,x')\!=\!\tfrac{d\Phi}{dx'}$. Consider a fixed frequency $\Omega$, which occupies a single column on the SLM. Realizing the axial group-velocity profile $\widetilde{v}(z)\!=\!\widetilde{v}_{1}+\tfrac{1}{m}\Delta\widetilde{v}(\tfrac{z}{L})^{m}$ requires that the local spatial frequency at $x'$ contributing to the on-axis ($x\!=\!0$) field at $z$ is $ck_{x}(\Omega,x')\!=\!\sqrt{2\omega_{\mathrm{o}}\Omega(1-\widetilde{n}(z))}$, where $z\!\approx\!\tfrac{k}{k_{x}}x'$ \cite{Motz21PRA}; see Fig.~\ref{Fig:ConceptAndSetup}(d). This provides an algorithm to establish the SLM phase $\Phi(\Omega,x')$. An approximation can be obtained when $\Delta\widetilde{v}\!<\!c$ and $\widetilde{n}_{1}\!\neq\!1$:
$k_{x}(\Omega,x')\!\approx\!k_{x1}(\Omega)+\Delta k_{x}(\Omega)(t\frac{x'}{x_{\mathrm{max}}'})^{m}$; where $k_{x1}(\Omega)\!=\!k_{x}(\Omega,0)$, $k_{x2}(\Omega)\!=\!k_{x}(\Omega,x_{\mathrm{max}}')$, $x_{\mathrm{max}}'\!\approx\!\tfrac{k_{x1}}{k_{\mathrm{o}}}L$, $\Delta k_{x}(\Omega)\!=\!k_{x2}(\Omega)-k_{x1}(\Omega)\!\approx\!\tfrac{k_{x1}(\Omega)c\Delta\widetilde{v}}{2m\widetilde{v}_{1}^{2}(1-\widetilde{n}_{1})}$, and $\Phi(\Omega,x')\!=\!\int_{0}^{x'}\!dx\,k_{x}(\Omega,x)$ is:
\begin{equation}
\Phi(\Omega,x')\!=\!k_{x1}(\Omega)x'+\frac{x_{\mathrm{max}}'\Delta k_{x}}{m+1}\left(\frac{x'}{x'_{\mathrm{max}}}\right)^{m+1},\,\, x'>0;
\end{equation}
a similar equation can be obtained for $x'\!<\!0$. Here $ck_{x1}(\Omega)\!=\!\sqrt{2\omega_{\mathrm{o}}\Omega(1-\widetilde{n}_{1})}$ corresponds to the initial group index $\widetilde{n}_{1}\!=\!\widetilde{n}(0)$, and $ck_{x2}(\Omega)\!=\!\sqrt{2\omega_{\mathrm{o}}\Omega(1-\widetilde{n}_{2})}$ to the terminal group index $\widetilde{n}_{2}\!=\!\widetilde{n}(L)$. We plot in Fig.~\ref{Fig:ConceptAndSetup}(f-j) the SLM phase $\Phi(\lambda,x')$ and its derivative $k_{x}(\lambda_{\mathrm{c}},x')\!=\!\tfrac{d\Phi}{dx'}$ at $\lambda_{\mathrm{c}}\!=\!799$~nm for $m\!=\!1$ [Fig.~\ref{Fig:ConceptAndSetup}(f)], $m\!=\!2$ [Fig.~\ref{Fig:ConceptAndSetup}(g)], $m\!=\!3$ [Fig.~\ref{Fig:ConceptAndSetup}(h)], $m\!=\!1.5$ [Fig.~\ref{Fig:ConceptAndSetup}(i)], and a composite acceleration profile comprising acceleration followed by deceleration such that the terminal group velocity at $z\!=\!L$ returns to its initial value at $z\!=\!0$.

Using this strategy, we realize the acceleration profiles depicted in Fig.~\ref{Fig:MeasuredAcceleration}. For each case, we plot the measured differential group delay obtained at
5-mm axial intervals (first row), the estimated local group velocity (second row), and acceleration (third row); $L\!=\!20$~mm throughout. We start with a propagation-invariant ST wave packet traveling at a fixed group velocity $\widetilde{v}(z)\!=\!\widetilde{v}_{1}\!=\!1.19c$ ($\theta\!=\!50^{\circ}$). The group delay measured in a reference frame moving at $\widetilde{v}_{1}$ is zero for all $z$, indicating a fixed $\widetilde{v}$ and $\widetilde{a}\!=\!0$ [Fig.~\ref{Fig:MeasuredAcceleration}(a)]. Next, we synthesize a ST wave packet with a group velocity $\widetilde{v}(z)\!\approx\!1.19c+0.2c(\tfrac{z}{L})$ that increases linearly with $z$ from $\widetilde{v}_{1}\!=\!1.19c$ to $\widetilde{v}_{2}\!=\!1.39c$, and thus constant acceleration $\widetilde{a}(z)L\!=\!\Delta\widetilde{v}\!=\!\widetilde{v}_{2}-\widetilde{v}_{1}\!=\!0.2c$ [Fig.~\ref{Fig:MeasuredAcceleration}(b)]. After being initially synchronized with the ST wave packet, the group delay measured in a reference frame moving at $\widetilde{v}_{1}$ is negative for all $z$, thereby indicating that the ST wave packet is speeding up. We also produced a decelerating ST wave packet with $\widetilde{v}(z)\!\approx\!1.19c-0.1c(\tfrac{z}{L})$ [Fig.~\ref{Fig:MeasuredAcceleration}(b)]. Here the group delay along $z$ is positive relative to the frame moving at $\widetilde{v}_{1}$, indicating that the ST wave packet is slowing down with respect to the reference.

The first two cases of constant-$\widetilde{v}$ [$m\!=\!0$; Fig.~\ref{Fig:MeasuredAcceleration}(a)] and constant-$\widetilde{a}$ [$m\!=\!1$; Fig.~\ref{Fig:MeasuredAcceleration}(b)] are in line with previous reports \cite{Clerici08OE,Lukner09OE,Chong10NP,Yessenov20PRL2}. We now proceed to describe our results for axially varying $\widetilde{a}(z)$. No such optical wave packets have been synthesized before to the best of our knowledge. First, we produce an accelerating ST wave packet with $\widetilde{v}(z)\!\approx\!1.19c+0.132c(\tfrac{z}{L})^{2}$ in which the acceleration increases linearly with $z$ [$m\!=\!2$; Fig.~\ref{Fig:MeasuredAcceleration}(c)]. As a second example, we synthesize a decelerating ST wave packet with $\widetilde{v}(z)\!\approx\!1.427c-0.12c(\tfrac{z}{L})^{3}$; i.e., it decelerates from an initial group velocity $\widetilde{v}_{1}\!=\!1.427c$ at a quadratic rate with $z$ [$m\!=\!3$; Fig.~\ref{Fig:MeasuredAcceleration}(d)]. Next, in Fig.~\ref{Fig:MeasuredAcceleration}(e) we plot measurements for a ST wave packet whose group velocity and acceleration vary according to a non-integer exponent of $z$. Specifically, $m\!=\!0.5$ and $\widetilde{v}(z)\!\approx\!1.19c+0.179c\sqrt{\tfrac{z}{L}}$, whereupon the wave-packet acceleration is $\widetilde{a}(z)L\!\approx\!0.089(\tfrac{z}{L})^{-0.5}$. The group velocity increases along $z$, but the associated acceleration rate decreases.

Finally, we show in Fig.~\ref{Fig:MeasuredAcceleration}(f) a ST wave packet with a composite acceleration profile in which $\widetilde{v}(z)\!\approx\!1.44c+0.45\tfrac{z}{L}-0.47c(\tfrac{z}{L})^{2}$. Starting off at $\widetilde{v}(0)\!=\!1.44c$, the wave packet first accelerates, but the rate of acceleration gradually decreases and reaches zero at $z\!\approx\!\tfrac{L}{2}\!=\!10$~mm, whereupon the wave packet reaches a maximum group velocity of $\widetilde{v}(\tfrac{L}{2})\!\approx\!1.55c$. Subsequently, the wave packet \textit{decelerates} and $\widetilde{v}$ drops until it returns to its initial value $\widetilde{v}(L)\!\approx\!\widetilde{v}(0)\!=\!1.44c$.  

After early emphasis on propagation-invariance as the main characteristic of ST wave packets \cite{Turunen10PO,FigueroaBook14}, recent work has produced counterparts in which a particular parameter is isolated and made to undergo prescribed axial dynamics while holding all the other parameters fixed; e.g., axial spectral encoding \cite{Motz21PRA}, pulse-width control by realizing arbitrary dispersion profiles in free space \cite{Yessenov21ACSP,Hall21PRA}, or periodic revivals of the spatio-temporal structure associated with discretized spectra (ST Talbot effects) \cite{Yessenov20PRL1,Hall21APLP,Hall21OL2}. Our strategy here (in addition to \cite{Yessenov20PRL2}) adds to this arsenal the possibility for arbitrary control over the axial distribution of the group velocity. It is important to emphasize that we are concerned with \textit{axial} acceleration of a wave packet in time, and not so-called \textit{transverse} acceleration of Airy \textit{beams} \cite{Siviloglou07OL,Siviloglou07PRL} where the peak of the monochromatic beam follows a parabolic trajectory in space as it freely propagates.

In conclusion, we have synthesized ST wave packets with fully controllable acceleration profiles in free space along the propagation axis. We have demonstrated for the first time a broad range of group-velocity and acceleration profiles, including: (1) constant-$\widetilde{v}$, $\widetilde{a}\!=\!0$; (2) linearly varying group velocity $\widetilde{v}\!\propto\!z$, or constant-$\widetilde{a}$; (3) quadratically varying group velocity $\widetilde{v}\!\propto\!z^{2}$, or linearly varying acceleration $\widetilde{a}\!\propto\!z$; (4) cubically varying group velocity $\widetilde{v}\!\propto\!z^{3}$, or quadratically varying acceleration $\widetilde{a}\!\propto\!z^{2}$; (5) fractional rates $\widetilde{v}\!\propto\!z^{0.5}$ and $\widetilde{a}\!\propto\!z^{-0.5}$; and (6) a composite acceleration profile in which the wave packet first accelerates from an initial group velocity to a maximum value, before the wave packet decelerates back to its initial group velocity. All these acceleration and deceleration profiles incorporate large variations in the group velocity $\sim~0.2c$ occurring over short axial distances $\sim\!20$~mm. By further extending our control over the axial propagation of ST wave packets, these results may pave the way to new sources of radiation, enable the acceleration of charged particles, and provide novel phase-matching schemes in nonlinear interactions in structured media. 

\section*{Funding}
U.S. Office of Naval Research (ONR) N00014-17-1-2458 and N00014-20-1-2789.

\section*{Disclosures}
The authors declare no conflicts of interest.

\bibliography{diffraction}

\end{document}